\documentclass[conference]{IEEEtran}
\IEEEoverridecommandlockouts

\usepackage[pdftex]{graphicx} 

\newlength\figureheight
\newlength\figurewidth
\usepackage{amsmath} 
\usepackage{amssymb} 
\usepackage{amsthm} 
\usepackage{thmtools} 
\usepackage{bm} 
\usepackage{csquotes} 
\usepackage{balance} 
\usepackage{cite} 
\usepackage{addlines} 

\usepackage[capitalize]{cleveref}
\crefname{equation}{\unskip}{\unskip}
\crefname{appsec}{Appendix}{Appendices}

\usepackage{xcolor}

\newcommand{\me}{\mathrm{e}}

\DeclareMathOperator*{\argmin}{arg\,min}

\newcommand{\x}[1][]{\bm{x}_{#1}}
\newcommand{\xh}[1][]{\hat{\bm{x}}_{#1}}
\newcommand{\xt}[1][]{\tilde{\bm{x}}_{#1}}

\newcommand{\trans}{\bm{F}}
\newcommand{\yt}[1][]{\tilde{\bm{y}}_{#1}}
\newcommand{\gain}[1][]{\bm{K}_{#1}}
\newcommand{\e}[1][]{\bm{e}_{#1}}
\newcommand{\mb}[1][]{\bar{m}_{#1}}

\newcommand{\No}{N_\mathsf{o}}
\newcommand{\Nw}{N_\mathsf{w}}
\newcommand{\Nv}{N_\mathsf{v}}
\newcommand{\Nb}{N_\mathsf{B}}
\newcommand{\Nc}{N_\mathsf{C}}
\newcommand{\Nk}{N_\mathsf{K}}

\renewcommand{\P}[1][]{\bm{P}_{#1}}
\newcommand{\Pt}[1][]{\tilde{\bm{P}}_{#1}}

\newcommand{\Q}{\bm{Q}} 
\newcommand{\R}{\bm{R}} 

\newcommand{\A}[1][]{\bm{A}^{(#1)}}
\newcommand{\B}[1][]{\bm{B}^{(#1)}}
\newcommand{\C}[1][]{\bm{C}^{(#1)}}
\newcommand{\Bx}{\bm{B}_{\xh[t]}}

\newcommand{\nc}{n_{\mathsf{C}}}
\newcommand{\nb}{n_{\mathsf{B}}}

\newcommand{\Bs}[1][]{\mathcal{B}^{(#1)}}
\newcommand{\Cs}[1][]{\mathcal{C}^{(#1)}}

\newcommand{\z}[1][]{\bm{z}_{#1}}
\newcommand{\obs}[1]{\bm{H}^{(#1)}}
\renewcommand{\H}[1]{\bm{H}}
\newcommand{\transpose}[1]{#1^{\textup{\textsf{T}}}}

\newcommand{\pn}{\bm{q}}

\newcommand{\on}{\bm{r}}

\title{Coded Distributed Tracking}

\author{
  \IEEEauthorblockN{
    Albin Severinson\IEEEauthorrefmark{2},
    Eirik Rosnes\IEEEauthorrefmark{2}, and
    Alexandre Graell i Amat\IEEEauthorrefmark{3}\IEEEauthorrefmark{2}
  }
  \vspace{0.1cm}

  \IEEEauthorblockA{
    \IEEEauthorrefmark{2}Simula UiB, Bergen, Norway}

  \IEEEauthorblockA{
    \IEEEauthorrefmark{3}Department of Electrical Engineering, Chalmers
    University of Technology, Gothenburg, Sweden}

 \thanks{The work of A.\ Graell i Amat was funded by the Swedish Research Council under grant
    2016-04253.}

}

\begin{document}
\maketitle
\begin{abstract}

  We consider the problem of tracking the state of a process that
  evolves over time in a distributed setting, with multiple observers
  each observing parts of the state, which is a fundamental
  information processing problem with a wide range of applications.
  We propose a cloud-assisted scheme where the tracking is performed
  over the cloud. In particular, to provide timely and accurate
  updates, and alleviate the straggler problem of cloud computing, we
  propose a coded distributed computing approach where coded
  observations are distributed over multiple workers. The proposed
  scheme is based on a coded version of the Kalman filter that
  operates on data encoded with an erasure correcting code, such that
  the state can be estimated from partial updates computed by a subset
  of the workers. We apply the proposed scheme to the problem of
  tracking multiple vehicles. We show that replication achieves
  significantly higher accuracy than the corresponding uncoded
  scheme. The use of maximum distance separable (MDS) codes further
  improves accuracy for larger update intervals. In both cases, the
  proposed scheme approaches the accuracy of an ideal centralized
  scheme when the update interval is large enough. Finally, we observe
  a trade-off between \emph{age-of-information} and estimation
  accuracy for MDS codes.

\end{abstract}

\section{Introduction}

Tracking the state of a process that evolves over time in a
distributed fashion is one of the most fundamental distributed
information processing problems, with applications in, e.g., signal
processing, control theory, robotics, and intelligent transportation
systems (ITS) \cite{Kalman60,Olfati07,Papadimitratos09}. These
applications typically require collecting data from multiple sources
that is analyzed and acted upon in real-time, e.g., to track vehicles
in ITS, and rely on timely status updates to operate effectively. The
analysis and design of schemes for providing timely updates has
received a significant interest in recent years. In a growing number
of works, timeliness is measured by the \emph{age-of-information}
(AoI) \cite{Yates12}, defined as the difference between the current
time, $t$, and the largest generation time of a received message,
$U(t)$, i.e., the AoI is $\Delta t = t - U(t)$.

Distributed tracking often entails highly demanding computational
tasks. For example, in many previous works the computational
complexity of the tasks performed by each node scales with the
cube of the state dimension, see, e.g., \cite{Mahmoud13,Olfati07} and
references therein. Thus, the proposed schemes are only suitable for
low-dimensional processes. A notable exception is the algorithm
proposed in \cite{Khan08}, where the overall process is split into
multiple overlapping subsystems to reduce the computational
complexity. However, the algorithm in \cite{Khan08} is based on
iterative message passing and potentially requires many iterations
to reach consensus, which makes it difficult to provide timely updates.

Offloading computations over the cloud is an appealing solution to
aggregate data and speed up demanding computations such that a
stringent deadline is met. In \cite{Kumar12}, a cloud-assisted
approach for autonomous driving was shown to significantly improve the
response time compared to traditional systems, where vehicles are not
connected to the cloud. However, servers in modern cloud computing
systems rarely have fixed roles. Instead, incoming tasks are
dynamically assigned to servers \cite{Verma2015}, which offers a high
level of flexibility but also introduces significant challenges. For
example, so-called \emph{straggling servers}, i.e., servers that
experience transient delays, may introduce significant delays
\cite{Dean2004}. Thus, for applications requiring very timely updates,
offloading over the cloud must be done with
care.

Recently, the use of erasure correcting codes has been proposed to
alleviate the straggler problem in distributed computing systems
\cite{Li2016,Lee2017,Severinson2018}. In these works, redundancy is
added to the computation such that the final output of the computation
can later be decoded from a subset of the computed results. Hence, the
delay is not limited by the slowest server.

In this paper, we consider a distributed tracking problem where
multiple observers each observe parts of the state of the system, and
their observations need to be aggregated to estimate the overall state
\cite{Olfati07,Khan08}. The goal is to provide timely and accurate
information about the state of a stochastic process. An example is
tracking vehicles to generate collision warning messages. We propose a
cloud-assisted scheme where the tracking is performed over the cloud,
which collects data from all observers. In particular, to speed up
computations, the proposed scheme borrows ideas from coded distributed
computing by distributing the observations over multiple workers, each
computing one or more partial estimates of the state of the system.
These partial estimates are finally merged at a monitor to produce an
estimate of the overall state.  To make the system robust against
straggling servers, which may significantly impair the accuracy of the
estimate unless accounted for, redundancy is introduced via the use of
erasure correcting codes. In particular, the observations are encoded
before they are distributed over the workers to increase the
probability that the information is propagated to the monitor. A
salient contribution of the paper is a coded filter based on the
Kalman filter \cite{Kalman60} that takes coded observations as its
input and returns a state estimate encoded with an erasure correcting
code.  Hence, the monitor can obtain an overall estimate from a subset
of the partial estimates via a decoding operation. We apply the
proposed scheme to the problem of tracking multiple vehicles using
repetition codes and random maximum distance separable (MDS) codes. We
show that replication achieves significantly higher accuracy than the
corresponding uncoded scheme and that MDS codes further improve
accuracy for larger update intervals. Notably, both schemes approach
the accuracy of an ideal centralized scheme for large enough update
intervals. Finally, for MDS codes we observe a trade-off between AoI
and accuracy, with update intervals shorter than some threshold
leading to significantly lower accuracy.

\section{System Model and Preliminaries}

We consider the problem of tracking the state of a stochastic process
over time in a distributed setting. The state at time step $t$
is represented by a real-valued vector $\bm{x}_t$ of length $d$ and
evolves over time according to
\begin{equation} \notag
  \bm{x}_t = \trans \bm{x}_{t-1} + \pn_t,
\end{equation}
where $\trans$ is the matrix representing the state transition model
and $\pn_t$ is a noise vector drawn from a zero-mean Gaussian
distribution with covariance matrix $\bm{Q}$. We denote by
$\hat{\bm{x}}_t$ the state estimate at time $t$ and we measure the
accuracy of the estimate by its root mean squared error (RMSE).

At each time step, a set of $\No$ observers,
$\mathcal{O} = \{ o_1, \dots, o_{\No} \}$, obtain noisy partial
observations of the state of the process. Specifically, the
observation made by observer $o$ at time $t$ is represented by the
vector
\begin{equation} \notag
\bm{z}_t^{(o)} = \obs{o} \bm{x}_t + \on^{(o)}_t,
\end{equation}
where $\obs{o}$ is a matrix of size $h^{(o)} \times d$ representing
the observation model of observer $o$ and $\on^{(o)}_t$ is a noise
vector drawn from a zero-mean Gaussian distribution with covariance
matrix $\bm{R}^{(o)}$. Furthermore, we denote by $\z[t]$ the overall
observation vector formed by concatenating the observations made by
all observers, $\z[t]^{(o_1)}, \dots, \z[t]^{(o_{\No})}$, and by $h$
the length of $\z[t]$. Similarly, we denote by $\bm{H}$ and $\on_t$
the overall observation model and noise vector, respectively, such
that $\z[t] = \bm{H} \bm{x}_t + \on_t$, and by $\bm{R}$ the covariance
matrix of $\on_t$. For simplicity we assume that all observations are
of equal dimension. We also assume that $h \geq d$ and that the
entries of each observation $\bm{z}_t^{(o)}$ are linear combinations
of a small number of entries of $\x[t]$, i.e., the observation
matrices $\obs{o}$ are sparse, as is the case, e.g., for an observer
measuring speed. The observations made by the $\No$ observers need to
be aggregated to estimate the overall state. Since $d$ may be large,
the work of aggregating the observations is performed in the cloud
over a set of $\Nw$ workers,
$\mathcal{W} = \{ w_1, \dots, w_{\Nw} \}$. We assume that the matrices
$\trans$, $\bm{Q}$, $\obs{o}$, and $\bm{R}^{(o)}$ are known.

\subsection{Probabilistic Runtime Model}
\label{sec:runtime_model}

We assume that workers become unavailable for a random time after
completing a computing task, which is captured by the exponential
random variable $V$ with probability density function
\cite{Li2016,Lee2017}
\begin{equation} \notag
  f_V(v) =
  \begin{cases}
    \frac{1}{\beta} \me^{-\frac{v}{\beta}} & v \geq 0 \\
    0 & v < 0
  \end{cases},
\end{equation}
where $\beta$ is used to scale the tail of the distribution, which
accounts for transient disturbances that are at the root of the
straggler problem. We refer to $\beta$ as the straggling parameter.

\subsection{Distributed Tracking}
\label{sec:distributed}

At time step $t$, each observer $o$ uploads its observation
$\bm{z}_t^{(o)}$ to the cloud, where the observations are encoded and
distributed over the $\Nw$ workers. Next, each worker $w$ that becomes
available during time step $t$ computes locally one or more partial
estimates of the state $\bm{x}_t$. These partial estimates are
forwarded to a monitor, which is responsible for computing the overall
estimate of $\x[t]$, denoted by $\xh[t]$, from the partial
estimates. Thus, the monitor has access to an updated state estimate
at the end of each time step, which can be used for other applications
(e.g., to generate collision warning messages in ITS). Finally, the
overall estimate is sent back to the workers to be used in the next
time step, i.e., we assume that all workers have access to $\xh[t-1]$
at time step $t$.

\subsection{Kalman Filter}
\label{sec:kalman}

Denote by $\xt[t]$ the prediction of the state at time step $t$ based
on the state estimate $\xh[t-1]$ at time step $t-1$ and the state
transition matrix $\trans$, i.e., $\xt[t] = \trans \xh[t-1]$, and by
$\Pt[t] = \trans \P[t-1] \transpose{\trans} + \Q$ the covariance
matrix of the error $\xt[t]-\x[t]$, where $\transpose{(\cdot)}$
denotes matrix transposition and $\P[t-1]$ is the covariance matrix of
the error $\xh[t-1] - \x[t-1]$ at time step $t-1$. The Kalman filter
is an algorithm for combining the predicted state $\xt[t]$ with an
observation $\z[t]^{(o)}=\obs{o}\x[t] + \on_t^{(o)}$ to produce an
updated state estimate $\xh[t]'$ with minimum mean squared error
\cite{Kalman60}. Let $\yt[t]^{(o)}= \z[t]^{(o)} - \bm{H}^{(o)} \xt[t]$
and denote by
$\bm{S}_t^{(o)} = \R^{(o)} + \bm{H}^{(o)} \Pt[t]
\transpose{\left(\bm{H}^{(o)}\right)}$ its covariance matrix. Then,
the updated state estimate is
$\xh[t]' = \xt[t] + \gain[t]^{(o)} \yt[t]^{(o)}$, where
$\gain[t]^{(o)} = \Pt[t] \transpose{\left(\bm{H}^{(o)} \right)}
\left(\bm{S}_t^{(o)} \right)^{-1}$ is the Kalman gain that determines
how the observation should influence the updated estimate.  The
covariance matrix of the error $\xh[t]' - \x[t]$ is
$\P[t]' = \left( \bm{I}_d - \gain[t]^{(o)} \bm{H}^{(o)} \right)
\Pt[t]$, where $\bm{I}_d$ is the $d \times d$ identity matrix. If more
than one observation is available, the estimate can be improved by
setting $\xt[t] \leftarrow \xh[t]'$ and $\Pt[t] \leftarrow \P[t]'$ and
repeating this procedure. After repeating this procedure for all
observations, the final estimate $\xh[t]$ is obtained.

\section{Proposed Coded Scheme}
\label{sec:scheme}

\begin{figure}[t]
  \centering
  \includegraphics[width=1\columnwidth]{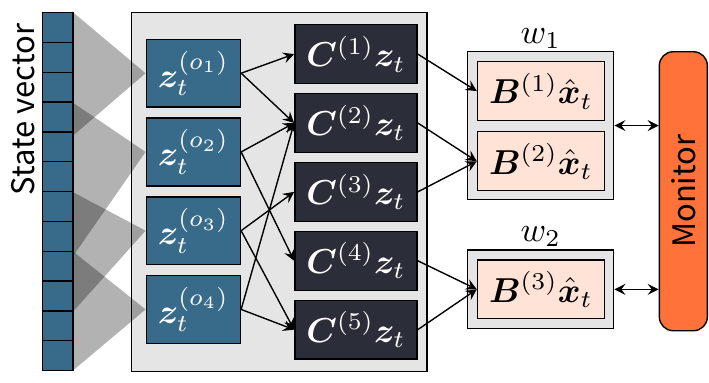}
  \vspace{-1ex}
  \caption{System overview example. Each of the $\No=4$ observers
    observes parts of the state, highlighted with a gray cone, and
    $\Nw=2$ workers compute $3$ coded state estimates from $5$ coded
    observations.}
  \label{fig:model}
  \vspace{-2ex}
\end{figure}

In this section, we introduce the proposed coded scheme. The key idea
is the use of two layers of coding to make the system robust against
straggling servers. The first layer consists of encoding the
observations and distributing them over multiple workers. More
specifically, the overall observation vector $\z[t]$ is encoded by an
$(\nc, h)$ linear erasure correcting code over the reals resulting in
the vector $\bm{C} \z[t]$, where $\bm{C}$ is a generator matrix of the
code. Next, the elements of $\bm{C} \z[t]$ are divided into $\Nc$
disjoint subvectors $\C[1]\z[t], \dots, \C[\Nc]\z[t]$, where $\C[i]$,
$i=1, \dots, \Nc$, is the corresponding division of the rows of
$\bm{C}$ into submatrices. We denote by $\nc^{(i)}$ the number of rows
of $\C[i]$. For the rest of the paper we refer to $\C[i] \z[t]$ as a
coded observation. This coding layer increases the probability that
the information from an observation propagates to the monitor in case
of delays.

The second layer of coding relates to the partial state estimates
computed by each worker. Specifically, we propose a coded version of
the Kalman filter that takes as its input the overall estimate at the
previous time step $\xh[t-1]$, provided by the monitor, one or more
coded observations $\C[i] \z[t]$ from the current time step, and
outputs a partial state estimate $\xh[t]^{(j)}$. The proposed filter
is such that the partial state estimate is equal to the estimate of
the regular Kalman filter multiplied by some matrix
$\B[j]$. Equivalently, it can be seen as an estimate of the state of a
process represented by the vector $\B[j] \x[t]$. Let $\bm{B}$ be a
generator matrix of an $(\nb, d)$ linear erasure correcting code over
the reals and let $\B[j]$ be a submatrix of $\bm{B}$ of size
$\nb^{(j)} \times d$ such that $\B[1], \dots, \B[\Nb]$ correspond to a
division of the rows of $\bm{B}$ into $\Nb \leq \Nc$ disjoint
submatrices. Hence, the partial estimates are symbols of the codeword
$\bm{B} \xh[t]$ and the monitor can recover the overall estimate
$\xh[t]$ from a subset of the partial estimates, ensuring that the
monitor has access to timely and accurate estimates even if multiple
workers experience delays.

Finally, we associate each coded state estimate $\B[j]\xh[t]$ with one
worker and each coded observation $\C[i]\z[t]$ with one coded state
estimate. We index the coded states associated with worker $w$ and the
coded observations associated with $\B[j] \xh[t]$ with the sets
$\Bs[w]$ and $\Cs[j]$, respectively. In total, worker $w$ receives the
set $\{\C[i] \z[t] : i \in \bigcup_{j\in \Bs[w]} \Cs[j] \}$ of
observations. The overall process is depicted in \cref{fig:model}.

\subsection{Coded Update}
\label{sec:codedupdate}

When a worker $w$ becomes available it first computes the set
$\{\B[j] \xh[t] : j \in \mathcal{B}^{(w)}\}$ of coded state estimates
associated with it. The worker also computes a randomly selected
subset of the Kalman gains associated with the uncoded filter, which
will later be used by the monitor to approximate the covariance matrix
$\P[t]$. Each coded state estimate $\B[j] \xh[t]$ is computed from the
previous state estimate $\xh[t-1]$ and the set
$\{\C[i]\z[t] : i \in \Cs[j] \}$ of coded observations associated with
it, and is computed independently from the other coded state estimates
using the following procedure. First, the worker computes
\begin{equation} \notag
  \xt[t]^{(j)} = \left( \B[j] \trans  \right) \xh[t-1]
\end{equation}
and the covariance matrix
\begin{equation} \notag
  \Pt[t]^{(j)} =
  \left( \B[j] \trans  \right)
  \P[t-1]
  \transpose{\left( \B[j] \trans \right)}
  + \B[j] \Q \transpose{\left( \B[j]  \right)}
\end{equation}
of the error $\xt[t]^{(j)} - \B[j]\x[t]$. Next, $\xt[t]^{(j)}$ and
$\Pt[t]^{(j)}$ are combined with the associated observations, i.e.,
the observations in $\{\C[i]\z[t] : i \in \Cs[j] \}$, one at a time,
to produce $\xh[t]^{(j)}$ and the covariance matrix $\P[t]^{(j)}$ of
the error $\xh[t]^{(j)} - \B[j] \x[t]$ in the following way. First,
consider a matrix $\A[i, j]$ such that
$\A[i, j] \B[j] = \C[i] \bm{H}$. Then,
\begin{equation} \notag
  \begin{split}
    \C[i] \z[t] & =
    \C[i] \left( \bm{H} \x[t] + \on_t \right) \\
    & = \C[i] \bm{H} \x[t]
    + \C[i] \on_t \\
    & = \A[i, j] \B[j] \x[t]
    + \C[i] \on_t,
  \end{split}
\end{equation}
i.e., the vector $\bm{C}^{(i)} \bm{z}_t$ can be considered as an
observation of the state $\B[j] \x[t]$ with observation matrix
$\A[i, j]$ and observation noise covariance matrix
$\C[i] \R \transpose{\left(\C[i] \right)}$. Hence, using an
observation $\C[i]\z[t]$, a partial coded state estimate
$\xh[t]'^{(j)}$ and the covariance matrix $\P[t]'^{(j)}$ of the error
$\xh[t]'^{(j)}-\B[j] \x[t]$ can be obtained as
\begin{equation} \label{eq:update}
  \xh[t]'^{(j)} = \xt[t]^{(j)} + \gain[t]^{(i, j)}  \yt[t]^{(i, j)},
\end{equation}
\begin{equation} \label{eq:update_p}
  \P[t]'^{(j)} = \left(\bm{I}_{\nb^{(j)}} -
    \gain[t]^{(i, j)} \A[i, j] \right) \Pt[t]^{(j)},
\end{equation}
where
\begin{equation} \notag 
  \yt[t]^{(i, j)} = \C[i] \z[t] - \A[i, j] \xt[t]^{(j)},
\end{equation}
\begin{equation} \notag 
  \gain[t]^{(i, j)}  = \Pt[t]^{(j)} \transpose{\left(
      \A[i, j] \right)} \left(\bm{S}_t^{(i, j)}
  \right)^{-1},
\end{equation}
\begin{equation} \notag 
  \bm{S}_t^{(i, j)}  = \C[i] \bm{R} \transpose{\left(\C[i] \right)} +
  \A[i, j] \Pt[t]^{(j)} \transpose{\left( \A[i, j] \right)}.
\end{equation}
Next, we let $\xt[t]^{(j)} \leftarrow \xh[t]'^{(j)}$ and
$\Pt[t]^{(j)} \leftarrow \P[t]'^{(j)}$ and repeat
\cref{eq:update,eq:update_p} for another observation until all
observations in $\{\C[i]\z[t] : i \in \Cs[j]\}$ have been used, at
which point the coded state estimate $\xh[t]^{(j)}$ has been
computed. The covariance matrix $\P[t]^{(j)}$ is only needed for
computing $\xh[t]^{(j)}$ and is discarded at this point. The worker
repeats the above procedure for each coded state estimate
$\xh[t]^{(j)}$, $j \in \Bs[w]$, assigned to it. Once finished, the
worker separately computes the Kalman gain of the uncoded filter
$\gain[t]^{(o)}$, as explained in \cref{sec:kalman}, associated with
some number $N_\mathsf{K}$ of observers $o$ selected uniformly at
random from $\mathcal{O}$. Finally, the coded state estimates are sent
to the monitor together with the $\gain[t]^{(i, j)}$ and
$\bm{S}_t^{(i, j)}$ matrices and the uncoded Kalman gains computed by the
worker, where they are used to recover the overall state estimate
$\xh[t]$ and the error covariance matrix $\P[t]$.

\subsection{Decoding}
\label{sec:decoding}

At the end of each time step $t$ the monitor attempts to recover
$\xh[t]$ from the partial coded state estimates $\xh[t]^{(j)}$
received from the workers. This corresponds to a decoding
operation. Denote by $\mathcal{U}_t$ the set of coded state estimates
the monitor receives at time step $t$ and by $\Bx$ the vertical
concatenation of the generator matrices associated with those
estimates. To decode, the monitor needs to solve for $\xh[t]$ in
$\Bx \xh[t] = \bm{y}_{\xh[t]}$, where $\bm{y}_{\xh[t]}$ is the
vertical concatenation of the vectors $\xh[t]^{(j)}$ in
$\mathcal{U}_t$. However, there are two issues that need to be
addressed before solving for $\xh[t]$. First, due to the dependence
structure of the tracking problem and the coding introduced, the
elements of $\bm{y}_{\xh[t]}$ will in general be correlated and have
varying variance, which must be accounted for to recover $\xh[t]$
optimally. Second, since the local estimates by the workers are noisy,
$\Bx \xh[t] = \bm{y}_{\xh[t]}$ typically does not have an exact
solution. We address the first issue by applying a so-called
\emph{whitening transform} to the original problem, i.e., we solve for
$\xh[t]$ in
$\bm{M}_{\xh[t]} \Bx \xh[t] = \bm{M}_{\xh[t]} \bm{y}_{\xh[t]}$, where
$\bm{M}_{\xh[t]}$ is a linear transform that has the effect of
uncorrelating and normalizing the variance of the elements of
$\bm{y}_{\xh[t]}$. The whitening transform $\bm{M}_{\xh[t]}$ is
computed from the singular value decomposition of the covariance
matrix of $\bm{y}_{\xh[t]}$, which we denote by $\bm{P}_{\xh[t]}$, as
in \cite{Kessy18}. The covariance matrix $\bm{P}_{\xh[t]}$ is given by
\cite[Eq. (6.47)]{Polavarapu04}, where the covariance matrix, Kalman
gain, observation model, and observation noise covariance matrix of
the uncoded filter update procedure are replaced by their coded
equivalents from \cref{sec:codedupdate}. We address the second issue
by finding the vector $\xh[t]$ that minimizes the $\ell_2$-norm of the
error, i.e., by solving
$\argmin_{\xh[t]} {\vert \vert \bm{M}_{\xh[t]} \Bx \xh[t] -
  \bm{M}_{\xh[t]} \bm{y}_{\xh[t]} \vert \vert}_2$. We achieve this by
decoding $\xh[t]$ using the LSMR algorithm \cite{Fong11}. The LSMR
algorithm is a numerical procedure for solving problems of this type
that takes an initial guess of the solution as its input and
iteratively improves on the solution until it has converged to within
some threshold. We give $\xt[t]$, which the monitor computes from
$\xh[t-1]$ as explained in \cref{sec:kalman}, as the initial guess
since the Euclidean distance between $\xh[t]$ and $\xt[t]$ typically
is small.

Next, the monitor approximates the error covariance matrix $\P[t]$
using the following heuristic. First, the monitor computes $\Pt[t]$
from $\P[t-1]$ as explained in \cref{sec:kalman}. Denote by $r$ the
maximum rank of $\bm{P}_{\xh[t]}$, i.e., the rank of $\bm{P}_{\xh[t]}$
when all workers are available, and by $r_{\xh[t]}$ the rank of the
given $\bm{P}_{\xh[t]}$. Now, if $r_{\xh[t]} < r$ we assume that the
monitor has insufficient information to recover $\xh[t]$ optimally and
we let $\P[t]=\Pt[t]$. On the other hand, if $r_{\xh[t]} = r$, we
assume that the monitor has recovered $\xh[t]$ optimally, and the
monitor computes $\P[t]$ from $\Pt[t]$ using the procedure for a full
update of the uncoded filter (see \cref{sec:kalman}). More formally,
denote by $\gain[t_o]^{(o)}$ the most recently received Kalman gain
corresponding to observer $o$ at time step $t_o$. Then, the monitor
computes
$\P[t]' = \left( \bm{I}_d - \gain[t_o]^{(o)} \bm{H}^{(o)} \right)
\Pt[t]$, assigns $\Pt[t]\leftarrow \P[t]'$, and repeats the procedure
for each remaining observer $o \in \mathcal{O}$. Finally, we let
$\P[t]\leftarrow \Pt[t]$.  Note that $\P[t]$ depends only on the
statistical properties of the observations, i.e., it can be computed
without access to the observations themselves.

\section{Design and Analysis of the Proposed Scheme}

We analyze the computational complexity of the proposed coded scheme,
design the generator matrices required for the coded filter update,
and choose how the computations are distributed over the workers.

\subsection{Computational Complexity}
\label{sec:processing_time}

We assume that the number of arithmetic operations performed by the
workers is dominated by the number of operations needed to invert
$\bm{S}$ when computing the Kalman gain, which requires in the order
of $n^3$ operations, where $n$ is the number of rows and columns of
$\bm{S}$. Since each worker computes $\Nk$ Kalman gains associated
with the uncoded filter in addition to those needed for the coded
state estimates, and due to our assumption that all uncoded
observations have equal dimension, the overall number of operations
performed by the workers can be approximated by
$\Nk \Nw \left(h^{(o)}\right)^3 + \sum_{w\in \mathcal{W}} \sum_{j \in
  \Bs[w]} \sum_{i\in \Cs[j]} \left( \nc^{(i)} \right)^3$.

\subsection{Code Design}
\label{sec:codedesign}

Here, we propose two strategies for designing the sets $\Bs[w]$ and
$\Cs[j]$ and the matrices $\B[1], \dots, \B[\Nb]$ and
$\C[1], \dots, \C[\Nc]$. The matrices $\A[i, j]$ are determined
implicitly since $\A[i, j] \B[j]=\C[i] \bm{H}$. The first design is
based on replication, which is a special case of MDS codes, whereas
the second is based on random MDS codes.

\subsubsection{Replication}

This design is based on replicating the tracking task at each worker,
i.e., the code rate is $h/\nc = 1/\Nw$. More formally, each worker
estimates $\x[t]$, i.e., $\Nb=\Nw$, $\B[j]=\bm{I}_d$,
$j=1, \dots, \Nb$, $\Bs[w_1]=\{1\}, \dots, \Bs[w_{\Nb}]=\{\Nb\}$, and
each state estimate is computed from the full set of observations,
i.e., $\Nc=\Nw \No$ and the observation encoding matrices and sets
$\Cs[j]$ associated with each estimate $\xh[t]^{(j)}=\xh[t]$ are such
that
$\{\C[i]\z[t] : i \in \Cs[j]\} = \{\z[t]^{(o)} : o \in
\mathcal{O}\}$. Note that the monitor can recover $\xh[t]$ and $\P[t]$
immediately upon receiving these values from any worker without
performing any additional computations. Hence, we let $\Nk=0$ and the
overall number of operations performed by the workers is approximately
$\frac{\No}{(h/\nc)} \left(h^{(o)}\right)^3$.

\subsubsection{Random MDS Coding}

This design is based on assigning a large number of coded state
estimates of dimension one to each worker, i.e., we let $\nb^{(j)}=1$,
$j=1, \dots, \Nb$. Furthermore, to ensure that the code is
\emph{well-conditioned}, i.e., the numerical precision lost due to the
coding is low, we generate $\bm{C}$ by drawing each element
independently at random from a standard Gaussian distribution
\cite{Chen05}. To satisfy the requirement
$\A[i, j] \B[j] = \C[i] \bm{H}$ we let $\B[j] = \C[i] \bm{H}$ and
$\A[i, j]=\bm{I}_1$. As a result, $\nc^{(i)}=1$, $i=1, \dots, \Nc$,
and we associate each observation one-to-one with a coded state
estimate, i.e., $\Nb=\Nc$ and $\Cs[1]=\{1\}, \dots,
\Cs[\Nb]=\{\Nb\}$. Next, we split the coded state estimates as evenly
as possible over the $\Nw$ workers, i.e., some workers are assigned
$\lfloor \Nb/\Nw \rfloor$ estimates and some are assigned
$\lceil \Nb/\Nw \rceil$ estimates.  Finally, we let
$\Nk = \left\lceil \frac{\No / (h/\nc)}{\Nw} \right\rceil$, which,
since $\nc^{(i)}=1$, $i=1, \dots, \Nc$, means that the overall number
of operations performed by the workers is approximately
$\frac{\No}{(h/\nc)} \left(h^{(o)}\right)^3$.

\section{Numerical Results}

To evaluate the performance of the proposed scheme, we consider a
distributed vehicle tracking scenario where $\Nw$ workers cooperate to
track the position of $\Nv$ vehicles $v_1, \dots, v_{\Nv}$ based on
observations received from the vehicles. We model the state of each
vehicle with a length-$4$ vector composed of its position and speed in
the longitudinal and latitudinal directions, i.e., the overall state
dimension is $d=4\Nv$. As in \cite{Soatti18}, we assume that the state
transition matrix of a single vehicle is
\begin{equation} \notag
  \bm{F}_\mathsf{v} =
  \begin{bmatrix}
    1 & 0 & \Delta t & 0 \\
    0 & 1 & 0 & \Delta t \\
    0 & 0 & 1 & 0 \\
    0 & 0 & 0 & 1
  \end{bmatrix}
\end{equation}
and that the associated covariance matrix is
\begin{equation} \notag
  \bm{Q}_\mathsf{v} = \bm{V}
  \begin{bmatrix}
    \sigma_\mathsf{a} & 0 \\
    0 & \sigma_\mathsf{a}
  \end{bmatrix}
  \transpose{\bm{V}},\;\; \text{with} \;\; \bm{V} =
  \begin{bmatrix}
    \Delta t^2 / 2 & 0 \\
    0 & \Delta t^2 / 2 \\
    \Delta t & 0 \\
    0 & \Delta t
  \end{bmatrix}.
\end{equation}
Hence, the combined state transition matrix and covariance matrix for
all vehicles is $\trans = \bm{I}_{\Nv} \otimes \bm{F}_\mathsf{v}$ and
$\Q = \bm{I}_{\Nv} \otimes \bm{Q}_\mathsf{v}$, respectively. We assume
that each vehicle observes its absolute position, e.g., using global
navigation satellite systems (GNSSs), and speed in the longitudinal
and latitudinal directions. The corresponding observation matrix is
$\bm{H}_\mathsf{v} = \bm{I}_{4}$, with associated covariance matrix
$\bm{R}_\mathsf{GNSS} = \text{diag}(\sigma_\mathsf{GNSS}^2,
\sigma_\mathsf{GNSS}^2, \sigma_\mathsf{speed}^2,
\sigma_\mathsf{speed}^2)$, where $\text{diag}(\cdot)$ denotes the
diagonal, or block-diagonal, matrix composed of the arguments of
$\text{diag}(\cdot)$ arranged along the diagonal.
Furthermore, similar to \cite{Soatti18} we assume that each vehicle
observes the distance and speed difference in the longitudinal and
latitudinal directions relative to a number $s < \Nv$ of other
vehicles using, e.g., radar or lidar. By combining these observations
in a cooperative manner the accuracy of the vehicle position estimates
can be improved compared to a system relying only on GNSS
observations. The covariance matrix associated with a relative
observation is
$\bm{R}_\mathsf{Rel.} = \text{diag}(\sigma_\mathsf{V2V}^2,
\sigma_\mathsf{V2V}^2, \sigma_\mathsf{speed}^2,
\sigma_\mathsf{speed}^2)$.

For each vehicle $v_i$, define the matrix $\bm{U}^{(v_i)}$ of size
$(s+1) \times \Nv$, where the first row corresponds to the absolute
observation of the vehicle and each of the $s$ remaining rows
correspond to an observation relative to another vehicle. The $i$-th
column of $\bm{U}^{(v_i)}$ has $s+1$ nonzero entries and the remaining
columns each have exactly one nonzero entry. For the first row of
$\bm{U}^{(v_i)}$ the $i$-th entry has value $1$, while the remaining
entries have value $0$. For each of the remaining rows the $i$-th
entry has value $-1$ and one other entry corresponding to the observed
vehicle has value $1$. For example, if $s=2$ and vehicle $v_i$ can
observe vehicles $v_j$ and $v_k$ the second and third row will have
value $1$ in column $j$ and $k$, respectively. Then, the observation
matrix for vehicle $v_i$ is
$\obs{v_i} = \bm{U}^{(v_i)} \otimes \bm{H}_\mathsf{v}$. The
corresponding observation noise covariance matrix is
$\text{diag}\left(\bm{R}_\mathsf{GNSS}, \bm{I}_s \otimes
  \bm{R}_\mathsf{Rel.}\right)$. Finally, $\bm{U}^{(v_i)}$ is generated
for one vehicle at a time such that the first vehicle observes
vehicles $v_2, \dots, v_{s+1}$, and, in general, vehicle $v_i$
observes vehicles $v_{(j \bmod \Nv) + 1}$, $j = i, \dots, i+s-1$.

We compare the performance of the proposed scheme with that of an
ideal centralized scheme where the monitor has unlimited processing
capacity and processes all observations itself using the procedure in
\cref{sec:kalman}. We also compare against the performance of an
uncoded scheme, where each observation is processed by a single worker
with no coding. More formally, we divide the $\No$ observations as
evenly as possible over the $\Nw$ workers, assigning
$\lfloor \No/\Nw \rfloor$ observations to some workers and
$\lceil \No/\Nw \rceil$ observations to the remaining workers. Next,
each worker estimates $\x[t]$ using the uncoded update procedure given
in \cref{sec:kalman}.  For this scheme, the monitor estimate is equal
to the average of the estimates received from the workers at each time
step, i.e., $\xh[t]$ is the average of the estimates in
$\mathcal{U}_t$ and $\P[t]$ is the average of the corresponding
covariance matrices.

We consider the vehicle tracking problem described above with
$\sigma_\mathsf{a}=0.3$, $\sigma_\mathsf{GNSS}=2$,
$\sigma_\mathsf{V2V}=0.5$, and $\sigma_\mathsf{speed}=10$. For all
schemes, we run $10$ simulations, each of $T=10000$ time steps, and
compute the RMSE of the position estimate at each time step. More
specifically, we denote by $\x[\mathsf{p}, t]$ and
$\xh[\mathsf{p}, t]$ the vectors composed of the entries of $\x[t]$
and $\xh[t]$ corresponding to position, e.g., entries $1, 2, 5, 6$ if
$\Nv=2$, and compute
$m_t \triangleq \sqrt{\frac{1}{d/2} \e[\mathsf{p}, t]
  \transpose{\e[\mathsf{p}, t]}}$, where
$\e[\mathsf{p}, t] = \xh[\mathsf{p}, t] - \x[\mathsf{p}, t]$, for
$t=1, \dots, T$. Next, for each simulation, to avoid any initial
transients, we discard the first $t_0-1$ samples
$m_1, \dots, m_{t_0-1}$. We let $t_0$ be the smallest value such that
\begin{equation} \notag
  \frac{
    \vert \mb[t_0:t_{\mathsf{m}}] -
    \mb[(t_{\mathsf{m}}+1):T] \vert
  }
  {
    \max\left(\mb[t_0:t_{\mathsf{m}}],
      \mb[(t_{\mathsf{m}}+1):T] \right)
  }
  \leq 0.1,
\end{equation}
where $t_{\mathsf{m}}=t_0 + \lfloor (T-t_0)/2 \rfloor$ and
$\mb[t_1:t_2]$ denotes the mean of $m_{t_1}, \dots,
m_{t_2}$. Finally, we plot the $90$-th percentile of the RMSE of the
position estimate over the concatenation of the remaining samples
from all simulations.

\begin{figure}[t]
  \centering
  \includegraphics[width=\columnwidth]{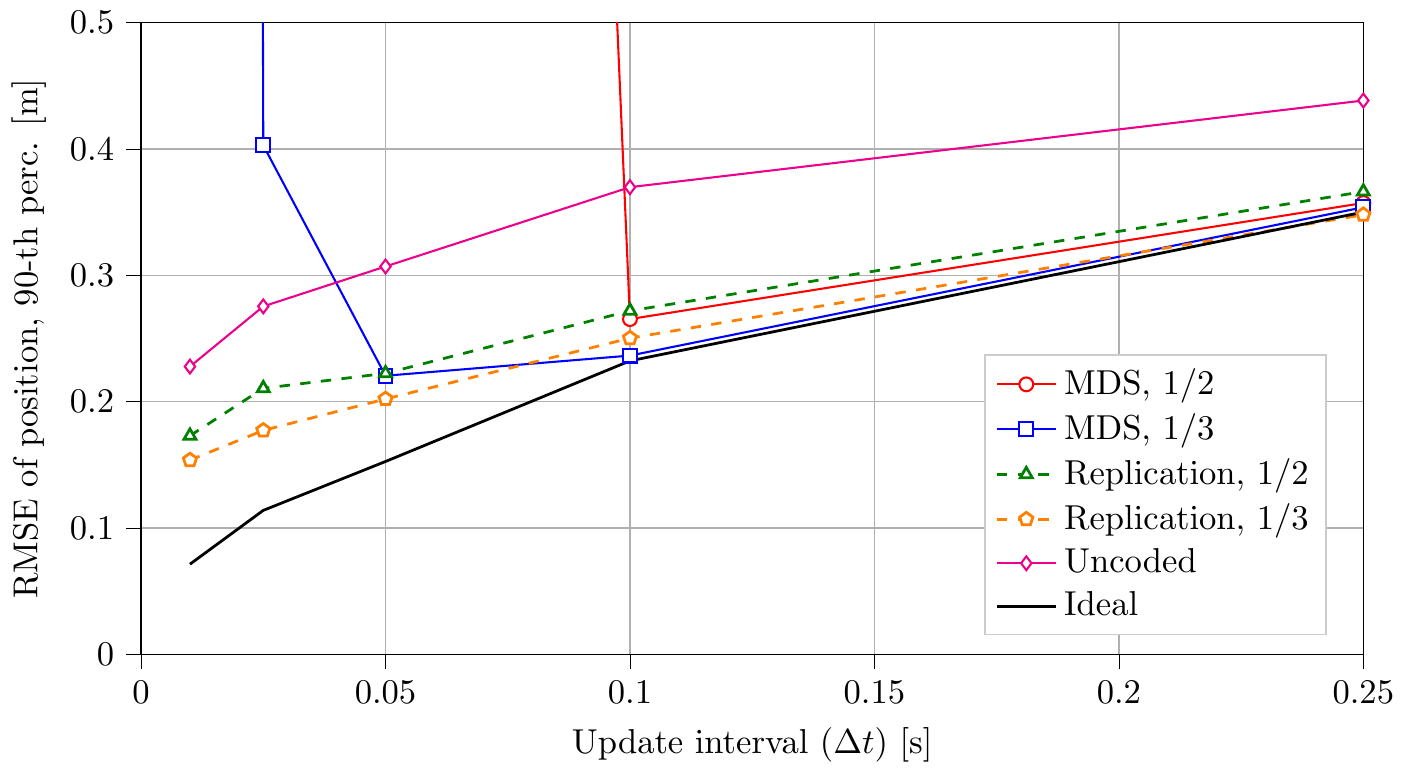}
  \vspace{-3ex}
  \caption{$90$-th percentile of the RMSE of the position over $10$
    simulations, each of $T=10000$ samples, as a function of
    $\Delta t$ for $\Nv=10$, $s=5$, $\Nw=16$ (for MDS codes), and
    $\beta=10$.}
  \label{fig:dt}
  \vspace{-1ex}
\end{figure}

In \cref{fig:dt}, we show the $90$-th percentile of the RMSE of the
position as a function of the update interval $\Delta t$ for
replication and random MDS codes with rates $1/2$ and $1/3$. For
replication the code rate is $h/\nc=1/\Nw$ (see
\cref{sec:codedesign}), i.e., the number of workers is $\Nw=2$ and
$\Nw=3$ for rates $1/2$ and $1/3$, respectively. MDS codes support an
arbitrary number of workers and we let $\Nw=16$ for this design. We
show the RMSE for $0.01 \leq \Delta t \leq 0.25$ since several
applications in ITS require an AoI in this range
\cite{Papadimitratos09}. There are $\Nv=10$ vehicles, each observing
$s=5$ other vehicles, and the straggling parameter is $\beta=10$,
i.e., workers become unavailable for $0.1$ seconds on average after a
filter update. Here, replication improves accuracy significantly
compared to the uncoded scheme, with a $90$-th percentile RMSE of
about $0.27$ and $0.25$ meters for code rates $1/2$ and $1/3$,
respectively, when $\Delta t=0.1$. MDS codes improve the accuracy
further at this point, with about a $2.4$\% and $5.5$\% smaller error
when compared at code rates $1/2$ and $1/3$, respectively. Finally,
for MDS codes we observe a trade-off between AoI and accuracy, with
update intervals shorter than some threshold, e.g., $\Delta t = 0.05$
for code rate $1/3$, leading to a higher RMSE since the probability of
the monitor collecting enough coded state estimates to decode $\xh[t]$
approaches zero when $\Delta t \to 0$.

\begin{figure}[t]
  \centering
  \vspace{-0ex}
  \includegraphics[width=\columnwidth]{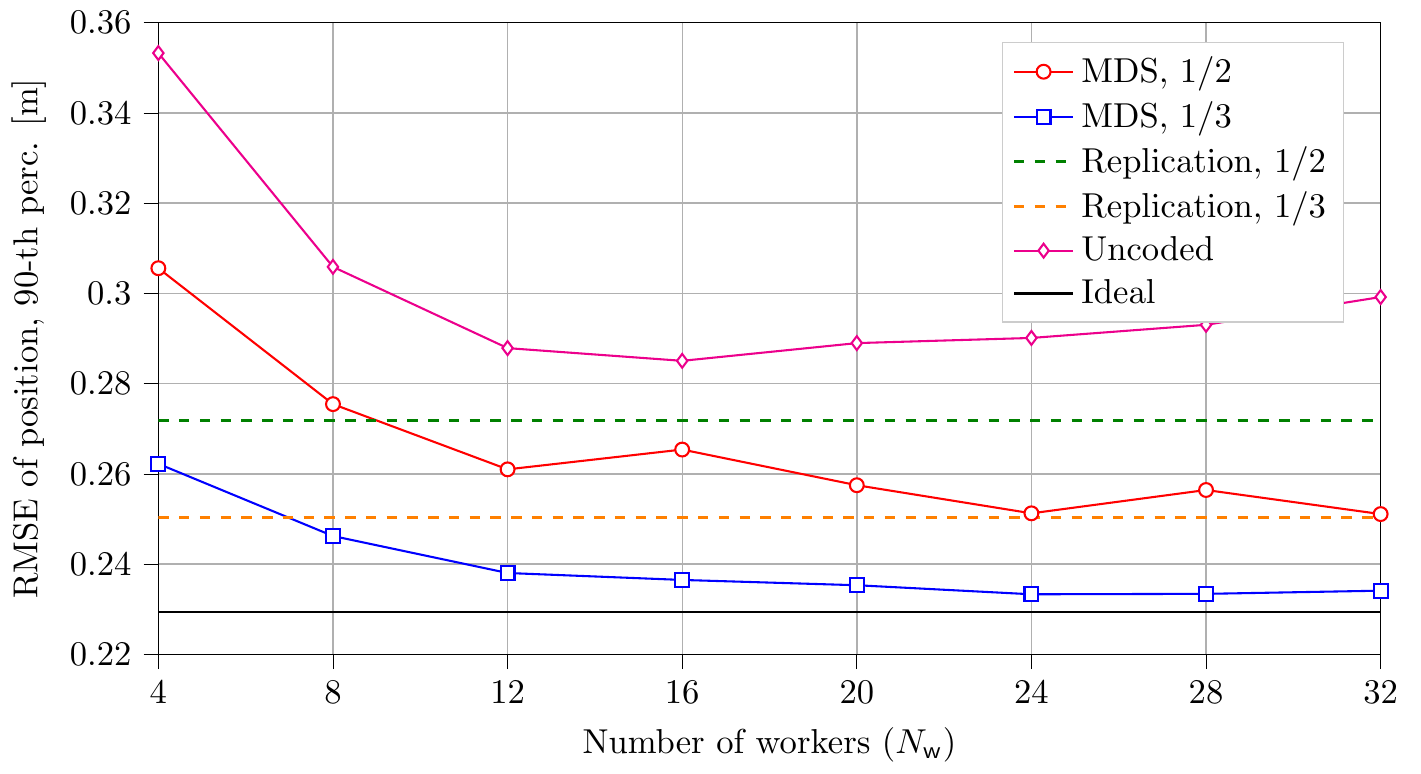}
  \vspace{-4ex}
  \caption{$90$-th percentile of the RMSE of the position over $10$
    simulations, each of $T=10000$ samples, as a function of $\Nw$ for
    $\Nv=10$, $s=5$, $\Delta t=0.1$, and $\beta=10$.}
  \label{fig:nv}
  \vspace{-3ex}
\end{figure}
In \cref{fig:nv}, we show the $90$-th percentile of the RMSE of the
position for random MDS codes as a function of the number of workers
$\Nw$ for $\Nv=10$ vehicles, each observing $s=5$ other vehicles,
$\Delta t=0.1$, and $\beta=10$. We also show the error of replication
(with $\Nw$ fixed to $2$ and $3$ for code rates $1/2$ and $1/3$,
respectively) and the uncoded and ideal schemes. The accuracy of the
design based on MDS codes generally improves with $\Nw$, since the
variance of the fraction of workers available in each time step
decreases. In some cases, e.g., for code rate $1/2$ and $\Nw=28$, the
error increases since the fraction of servers needed to decode
$\xh[t]$ may increase if the number of coded state estimates does not
divide evenly over the workers. Here, the error of MDS codes is lower
than that of replication when $\Nw \geq 12$ and $\Nw \geq 8$ for code
rates $1/2$ and $1/3$, respectively. We also observe that the
performance does not improve significantly beyond some number of
workers.

\section{Conclusion}
\balance

We presented a novel scheme for tracking the state of a process in a
distributed setting, which we refer to as coded distributed
tracking. The proposed scheme extends the idea of coded distributed
computing to the tracking problem by considering a coded version of
the Kalman filter, where observations are encoded and distributed over
multiple workers, each computing partial state estimates encoded with
an erasure correcting code, which alleviates the straggler problem
since missing results can be compensated for. The proposed coded
schemes achieves significantly higher accuracy than the uncoded scheme
and approaches the accuracy of an ideal centralized scheme when the
update interval is large enough. We believe that coded distributed
tracking can be a powerful alternative to previously proposed
approaches.

\section*{Acknowledgment}
The authors would like to thank Prof.\ Henk Wymeersch for fruitful
discussions and insightful comments.

\balance 
\bibliographystyle{IEEEtran}
\bibliography{manuscript}{}

\end{document}